\documentclass[conference]{IEEEtran}

\usepackage{graphics}
\usepackage{epstopdf}
\usepackage[pdftex]{graphicx}
\usepackage{stfloats}
\usepackage{array}
\usepackage{float}
\usepackage[cmex10]{amsmath}
\usepackage{amsmath,amsfonts,amssymb}
\usepackage{graphicx,graphics}
\usepackage{enumerate}
\usepackage{color}
\usepackage{verbatim}
\usepackage{amsthm}
\usepackage{multirow}
\usepackage{subcaption}
\usepackage{siunitx}
\usepackage{amsmath,amsthm}
\usepackage{cleveref}
\usepackage[font=footnotesize]{caption}
\usepackage{fixltx2e}
\usepackage[USenglish]{babel}

\begin{document}


\title{Impulsive Noise Detection in OFDM-based Systems: A Deep Learning Perspective}

\author{
Reza Barazideh$^{\dag}$, Solmaz Niknam$^{\dag}$, and Balasubramaniam Natarajan$^{\dag}$\\
\small $^{\dag}$ Department of Electrical and Computer Engineering\\
 Kansas State University, Manhattan, KS, USA.\\
Email:\{rezabarazideh, slmzniknam, bala\}@ksu.edu}


\maketitle

\begin{abstract}
Efficient removal of impulsive noise (IN) from received signal is essential in many communication applications. In this paper, we propose a two stage IN mitigation approach for orthogonal frequency-division multiplexing (OFDM)-based communication systems. In the first stage, a deep neural network (DNN) is used to detect the instances of impulsivity. Then, the detected IN is blanked in the suppression stage to alleviate the harmful effects of outliers. Simulation results demonstrate the superior bit error rate (BER) performance of this approach relative to classic approaches such as blanking and clipping that use threshold to detect the IN. We demonstrate the robustness of the DNN-based approach under (i) mismatch between IN models considered for training and testing, and (ii) bursty impulsive environment when the receiver is empowered with interleaving techniques.
\end{abstract}


\section{Introduction}

Impulsive noise (IN) can significantly degrade the performance of any communication system. Although orthogonal frequency-division multiplexing (OFDM) is inherently more resistant to IN than single carrier modulation, system performance can still degrade if IN power exceeds a certain threshold and its effect gets spread over all subcarriers \cite{ghosh1996}. Many techniques have been explored in prior efforts to mitigate the effect of IN. For example, the high amplitude and the short duration of IN are considered as the main features for threshold based IN mitigation approaches. Memoryless nonlinear approaches such as clipping \cite{Tseng-2012-robust-clipping}, blanking \cite{Blanking}, and multiple-threshold blanking/clipping \cite{MultiThershold_Rozic_2018} are the most common methods in this category. In \cite{AdaptiveNoiseMitigation-2010}, a threshold optimization based on Neyman-Pearson criterion is proposed and an analytical equation for the quasi-optimal blanking and clipping thresholds is provided in \cite{DesignBLNCLP_Oh_2017}. Authors in \cite{Khodam_Latincom,Khodam_ICC,Khodam_TVT,Khodam_Ocean} take advantage of analog domain processing where the impulsive noise is still broadband and distinguishable. Note that, determining thresholds in analog domain techniques is not trivial. The performance of threshold based nonlinear approaches is highly sensitive to the selected thresholds and as shown in \cite{Zhidkovn08_Simpleanalysis}, the performance of all these methods degrades dramatically in severe impulsive environment.

Machine learning methods such as deep learning are becoming popular in growing number of applications in signal and image processing~\cite{Kaliraj_efficientapproach_2010, Kauppinen_Speech_2002}, and resource allocation in wireless networks~\cite{Rouhi_PWA_HetNet_2018,Rouhi_JPWA_2018}. If appropriate network structures and processing strategies are employed, deep neural network (DNN) may be used as powerful tools for efficient detection of impulse noise because of their ability to learn from examples and capability to account for uncertainty that is common in the most communication applications. Additionally, in classical outlier detection approaches, determining the optimum threshold is the main challenge as this threshold will vary in response to channel conditions and model mismatches. Lastly, the high peak-to average-power-ratio (PAPR) of OFDM signals can also degrade the performance of the classical methods. As always, there is a compromise between detection and false alarm probability in the traditional threshold based methods.

To overcome the aforementioned drawbacks, we propose a machine learning based IN suppression strategy for an OFDM-based communication system. The proposed IN mitigation approach comprises of two stages: (i) IN detection and (ii) IN suppression. In the first stage, a DNN is used to detect the IN corrupted signal instances. Then, the detected IN can be either blanked or clipped in the suppression stage to alleviate the harmful effects of outliers. The proposed DNN-based IN detection approach can be used in conjunction with any IN mitigation strategy as the operation of the detector is completely independent of the noise removal operator. The proposed DNN consists of multiple layers (input, hidden, output) with nodes in a fully connected structure that maps input data into appropriate outputs. Each node in the hidden layers has a nonlinear activation function which helps to distinguish data that are not linearly separable. Here, the DNN uses the current sample value, median deviations filter output~\cite{Kong_MedianDeviation_1998}, and Rank-Ordered Absolute Differences (ROAD) statistic~\cite{Garnett_ROAD_2005} as the inputs to determine if the current sample is corrupted by IN or not. Bit error rate (BER) performance in an OFDM-based communication system is used to evaluate and compare the capability of the proposed DNN-based IN mitigation approach with other conventional approaches such as blanking and clipping. The robustness of the proposed approach is highlighted by testing the performance with IN model different from the model used for training. In addition, we evaluate the robustness of our method in bursty IN when the receiver is accompanied by time domain interleaving techniques. Simulation results show that the DNN-based approach offers up to 2 dB gains relative to blanking and clipping at BER $10^{-3}$.


The remainder of this paper is organized as follows. Section~\ref{sec:System and Noise Model} describes the system and noise models. Section~\ref{sec:DNN} presents the structure of the proposed DNN and its input features. The proposed algorithm for IN mitigation is detailed in Section~\ref{sec:Imp Mitigation}. The performance of the IN detector is analyzed in Section~\ref{sec:Simulation results} and finally conclusions are drawn in Section~\ref{sec:Conclusion}.

\section{System Model}\label{sec:System and Noise Model}

%

\begin{figure*}
\centering
\includegraphics[scale=.5]{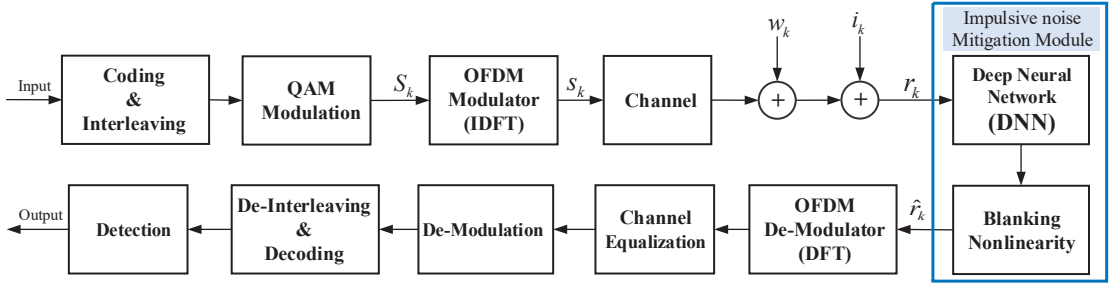}
\caption{System model block diagram.}
\label{fig:System Model}
\end{figure*}

Consider the OFDM system shown in Fig.~\ref{fig:System Model}. At the transmitter, information bits are channel coded and then the encoded bits are interleaved. Subsequently, the interleaved data is modulated and then passed through an inverse discrete Fourier transform (IDFT) module to generate OFDM symbols over orthogonal subcarriers. In general, an OFDM symbol can be constructed with $M$ non-data subcarriers and $N-M$ data subcariers. The non-data subcarriers are either pilots for channel estimation and synchronization, or nulled for spectral shaping and ICI reduction. Let the nonoverlapping sets of data, pilot, and null subcarriers be defined as $S_D$, $S_P$, and $S_N$, respectively. Therefore, after digital-to-analog conversion the transmitted signal envelope in the time domain can be expressed as
\begin{equation}
s(t) = \frac{1}{{\sqrt N }}\sum\limits_{k \in {S_A}} {{S_k}\,\,{{\rm{e}}^{j\frac{{2\pi kt}}{T_s}}}} ,{\mkern 1mu} {\mkern 1mu} {\mkern 1mu} {\mkern 1mu} {\mkern 1mu} 0 < t < T_s,
\end{equation}
where $S_A=S_D\cup S_P$ represents the set of active subcarriers; $S_k$ is the modulated symbol on the $k^{\rm{th}}$ subcarrier; and $T_s$ is the OFDM symbol duration.
The channel can be modeled as a linear time-varying system described by the channel impulse response
\begin{equation}\label{eq:Channel_Response}
c(\tau ) = \sum\limits_{p = 1}^L {{b_p}{\mkern 1mu} \delta (\tau  - {\tau _p})} ,
\end{equation}
where $L$ is the length of the channel impulse response; $b_p$ and $\tau_p$ are the amplitude and the delay of the $p^{\rm{th}}$ multipath component, respectively. Therefore, the received signal after down conversion, analog-to-digital conversion, guard interval removing, and synchronization can be expressed as
\begin{equation}\label{eq:Received_Signal}
{r_k} = \sum\limits_{p = 1}^L {{b_p}{s_{k - \tau_p}}}  + {n_k} \,,\,\,\,\,\,k = 0,\,1,\,.\,.\,.,\,N - 1
\end{equation}
where $s_k=s(kT_s/N)$; $n_k={w_k} + {i_k}$ is the mixture of additive white Gaussian noise (AWGN) $w_k$ and IN $i_k$.

%

Here it is assumed that the noise samples $n_k$ are uncorrelated and their distribution can be expressed in terms of multi-component mixture-Gaussian model \cite{Zhidkovn08_Simpleanalysis}. Corresponding to this model, the probability density function (PDF) of the noise samples $n_k$ is obtained as
\begin{equation}\label{eq:Imp_Model}
P({n_k}) = \sum\limits_{j = 0}^{J-1} {{p_j}G({n_k}\left| {\sigma _j^2} \right.)}
\end{equation}
where $G\left( {{n_k}\left| {{\sigma ^2}} \right.} \right)$ is the PDF of the complex Gaussian variable with zero-mean and variance $\sigma^2$, and $\left\{ {{\sigma _0},{\sigma _1},...,{\sigma _{J - 1}}} \right\}$ and $\left\{ {{p _0},{p _1},...,{p _{J - 1}}} \right\}$ are the model parameters such that $\sum\limits_{j = 0}^{J - 1} {{p_j} = 1}$. The noise model \eqref{eq:Imp_Model} can support two commonly used IN models. The first IN model is a two component mixture-Gaussian noise model or Bernoulli Gaussian (BG) noise model \cite{ghosh1996} with model parameters corresponding to
\begin{equation}\label{eq:BG_Model}
J = 2,\,\,\,{p_0} = 1 - \epsilon ,\,\,\,{p_1} = \epsilon ,\,\,\,\sigma _0^2 = \sigma _w^2,\,\,\,\sigma _1^2 = \sigma _w^2 + \sigma _i^2.
\end{equation}
Here $\epsilon$ is the probability of the incoming impulse noise, $\sigma _w^2$ is the variance of AWGN component, and $\sigma _i^2$ presents the variance of the IN. The expression in \eqref{eq:Imp_Model} can also be used to characterize a Middleton Class A (MCA) IN model~\cite{Middleton_ClassA_1983} with the following parameters
\begin{equation}\label{eq:MCA}
J = \infty ,\,\,\,{p_j} = \frac{{{{\rm{e}}^{ - A}}{A^j}}}{{j!}},\,\,\,\sigma _j^2 = \frac{{j{A^{ - 1}} + \Gamma }}{{1 + \Gamma }}\sigma _n^2,\,\,\,j = 0,1,...,\infty \end{equation}
where $\sigma _n^2$ is the noise variance of $n_k$, $A$ is the impulsiveness index designed as the product of the mean number of impulses per time unit and the mean length of an impulse (in time units), and $\Gamma= \sigma _w^2/\sigma _i^2$ denotes the background-to-IN power ratio~\cite{Middleton_ClassA_1983}.
The noise model in \eqref{eq:Imp_Model} is used to train the proposed DNN. In order to investigate the system performance when there is a model mismatch between training and testing, we also consider Symmetric Alpha Stable (S$\alpha$S) IN which can be expressed as \cite{Nikias_SaS_1996}
\begin{equation}\label{eq:SaS}
{n_k} \sim S\left( {\alpha ,\beta ,\gamma ,\mu } \right)
\end{equation}
where $\alpha  \in (0,2]$ denotes the stability parameter that sets the degree of the impulsiveness of the distribution; $\mu  \in \mathbb{R}$ is the location parameter; $\beta \in  [-1, 1]$ is called the skewness parameter and is a measure of asymmetry ($\beta = 0$ for S$\alpha$S distribution); $\gamma  \in (0,\infty)$ represents the scale parameter which is a measure of the width of the distribution.

\section{Deep Neural Network Design}\label{sec:DNN}

In order to deal with IN, a DNN is exploited to find the instances of impulsivity. DNN is a black-box approach that can be used to model any nonlinear system if properly trained. In this section, the structure of DNN is introduced and then the input features are presented.

\subsection{DNN Structure}\label{subsec:DNN_Structure}
\begin{figure}
\centering
\includegraphics[width=.45\textwidth,height=50mm]{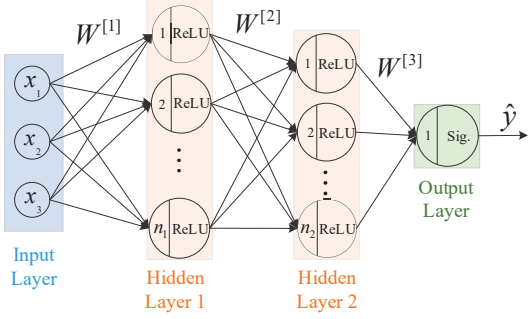}
\caption{Block diagram of the DNN.}
\label{fig:DNN}
\end{figure}

As shown in Fig.~\ref{fig:DNN}, the considered neural network consists of two hidden layers with $n_1$ and $n_2$ hidden neurons in each layer, respectively. Typically, there is no analytical method to choose the number of layers and neurons, and hence they are determined experimentally on a trial and error basis. Here, $\textbf{x}=[x_1,x_2,x_3]^T$ represents the input vector consisting of three features (as discussed in the next subsection) and $\hat y$ denotes the output of the DNN. There is only one node in the output layer, which generate a binary sequence of zeros and ones. Note that the soft outputs of DNN will be rounded off to a $0$ or $1$. An output $1$ indicates that the received sample $r_k$ is corrupted by IN and output $0$ implies that the $k^{th}$ received sample is uncorrupted. According to~Fig.~\ref{fig:DNN}, the relation between layers can be expressed as
\begin{align} \notag
\textbf{A}^{^{[1]}} &= {g^{[1]}}\left( {{\textbf{W}^{[1]}}\textbf{x} + {\textbf{b}^{[1]}}} \right)\\ \notag
\textbf{A}^{^{[2]}} &= {g^{(2)}}\left( {{\textbf{W}^{[2]}}{\textbf{A}^{(1)}} + {\textbf{b}^{[2]}}} \right)\\
{\hat y} &= {g^{[3]}}\left( {{\textbf{W}^{[3]}}{\textbf{A}^{[2]}} + {\textbf{b}^{[3]}}} \right),
\end{align}
where $\textbf{W}^{[l]}$, $\textbf{b}^{[l]}$, and $g^{[l]}$ are the parameter matrix, bias vector, and activation function of $l^{th}$ layer that will be applied to the output of the previous layer. The activation function is a nonlinear function in general, but can also be designed to retain linearity in the transformation process. In this paper, the Rectified linear unit (ReLU) function is used for the hidden layers and a Sigmoid function is used in the output layer. The ReLU and Sigmoid functions are expressed as
\begin{equation}\label{eq:Relu}
{\rm{ReLU}}(x)={\rm{max}}(x,0),
\end{equation}
\begin{equation}\label{eq:Sigmoid}
{\rm{Sigmoid}}(x)=\frac{1}{{1 + {{\rm{e}}^{ - x}}}}.
\end{equation}
Loss or cost function is a function that returns the loss or penalty associated with a predicted value $\hat y$ when the true value is $y$ over the entire training set. This loss function value decreases when the difference between the predicted value and the correct value decreases. The loss function that is used in this work corresponds to
\begin{align}\notag
\mathcal{L} (\textbf{W},\textbf{b}) =&  - \frac{1}{m}\left[ {\sum\limits_{i = 1}^m {{y_i}\log ({{\hat y}_i}) + (1 - {y_i})\log (1 - {{\hat y}_i})} } \right] \\
 &+ \frac{\lambda }{{2m}}\sum\limits_{l = 1}^{L - 1} {\sum\limits_{i = 1}^{{n_l}} {\sum\limits_{j = 1}^{{n_{l + 1}}} {W_{ij}^2} } },
\end{align}
where $m$ is the number of training samples; $n_l$ represents the number of neurons in layer $l$; and $\lambda$ denotes the regularization hyper parameter that is used to prevent over-fitting in the training phase. The DNN aims to determine the weights $\textbf{W}$ and the bias vector $\textbf{b}$ that minimize the loss function, i.e., 
\begin{equation}
\mathop {\min }\limits_{\textbf{W},\textbf{b}} \,\mathcal{L}(\textbf{w},\textbf{b}).
\end{equation}
The proposed DNN is trained using the back-propagation algorithm along with Adam optimization algorithm \cite{Adam_Optimizer}.  The Adam optimization is an extension to stochastic gradient descent and has recently seen broader adoption in deep learning applications. Adam computes adaptive learning rates for each parameter $\Theta$ at time instant $k$. According to the Adam algorithm, the update rule for each parameter $\Theta$ in layer $l$ is given by
\begin{equation}
\Theta _{k + 1}^{[l]} = \Theta _k^{[l]} - \frac{\eta }{{\sqrt {\hat \upsilon _k^{[l]}}  + \varepsilon }}\hat m_k^{[l]}.
\end{equation}
Here, $\eta$ is learning rate hyper parameter and
\begin{align}
m_k^{[l]} &= {\beta _1}m_{k - 1}^{[l]} + (1 - {\beta _1})\frac{{\partial \mathcal{L}(\Theta )}}{{\partial {\Theta ^{[l]}}}}\\ \notag
\hat m_k^{[l]} &= \frac{{m_k^{[l]}}}{{1 - \beta _1^k}},
\end{align}
\begin{align}
\upsilon _k^{[l]} &= {\beta _2}\upsilon _{k - 1}^{[l]} + (1 - {\beta _2}){\left( {\frac{{\partial \mathcal{L}(\Theta )}}{{\partial {\Theta ^{[l]}}}}} \right)^2} \\ \notag
\hat \upsilon _k^{[l]} &= \frac{{\upsilon _k^{[l]}}}{{1 - \beta _2^k}},
\end{align}
where the proposed default values are $\beta_1=0.9$, $\beta_2=0.999$, $\varepsilon = 10^{-8}$, and the initial value for $m_0^{[l]}$ and $\upsilon _0^{[l]}$ are randomly chosen.

\subsection{DNN Input Features}

Feature extraction is one of the most important aspects of machine learning because it turns raw data into information that is suitable for inferencing. Feature extraction eliminates the redundancy present in many types of measured data, facilitating generalization which is critical to avoiding over-fitting during the learning phase. According to Fig.~\ref{fig:DNN}, the input layer has three nodes which are (i) the current sample value, (ii) Rank-Ordered Absolute Differences (ROAD) statistic, and (iii) median deviations filter output. In the following we briefly introduce the ROAD and median deviation features.

\subsubsection{ROAD Value}

The ROAD value is an efficient statistic for distinguishing between corrupted and uncorrupted samples as its value is high for noisy samples and low for uncorrupted samples \cite{Garnett_ROAD_2005}. In general, ROAD factor is widely used in image processing for two dimensional (2D) signals. Here, we compute the ROAD factor for a one dimensional received signal as follows:

i. The absolute difference between the centre sample and the remaining samples of a $(1\times 2n)$ vector is calculated and denoted by ${\textbf{d}_{(k)}}$ which consists of $2n$ elements:
\begin{equation}
{\textbf{d}_{(k)}} = \left| {{r_k} - \left[ {{r_{k - n}},...,{r_{k - 1}},{r_{k + 1}},...,{r_{k + n}}} \right]} \right|
\end{equation}

ii. Sort $\textbf{d}_{(k)}$ values in increasing order:
\begin{equation}
\textbf{b}_{(k)}={\rm{sort}}({\textbf{d}_{(k)}})
\end{equation}

iii. The ROAD factor is calculated by summing up the first $n$ values of $\textbf{b}_{(k)}$:
\begin{equation}
{\rm{ROAD}} = \sum\limits_{k = 1}^n {{\textbf{b}_{(k)}}}.
\end{equation}

\subsubsection{Median Deviations Filter}

The median-deviations filter to obtain $e_k$ can be expressed as
\begin{equation}\label{eq:Median}
e_k = {r_k} - {\rm{median}}\left( {\left[ {{r_{k - n}},...,{r_k},...,{r_{k + n}}} \right]} \right),
\end{equation}
where the median filter used in \eqref{eq:Median} is a standard median filter which operates on a moving window of $2n+1$ samples.

\section{Impulsive Noise Mitigation}\label{sec:Imp Mitigation}

After the proposed DNN determines if a received sample is contaminated with IN or not, a simple memoryless nonlinear preprocessor such as blanking can be used to alleviate the effect of IN. Therefore, the output of blanking nonlinearity can be expressed as
\begin{equation}\label{eq:Blanking}
{{\hat r}_k} = \left\{ \begin{array}{l}
{r_k},\,\,\,\,\,\,\,\,\,\,{{\hat y}_k} = 0\\
0,\,\,\,\,\,\,\,\,\,\,\,\,{{\hat y}_k} = 1
\end{array} \right.,
\end{equation}
where ${{\hat y}_k}$ is the output of the DNN. It is worth mentioning that one can use other nonlinear preprocessors proposed in the literature to suppress the impact of IN. This extension is straightforward and is not the main focus of this paper. After IN mitigation a discrete Fourier transform (DFT) module is used to transform the time domain signal to the frequency domain. The DFT module is followed by frequency domain equalization that depends on channel estimation which can be performed based on pilot subcarriers. Viterbi soft decoding is used to decode the demodulated signal and then detection is performed based on the modulation scheme used.

\section{Simulation results}\label{sec:Simulation results}

In this section, an OFDM-based communication system with QPSK modulation in the presence of channel fading, channel coding, and IN is studied. The BER performance is used to compare the proposed DNN-based IN mitigation with other conventional approaches such as blanking and clipping. Since the distribution of the received OFDM signal in case of no IN can be considered as Gaussian, the threshold value for blanking and clipping in all scenarios is obtained based on the approach provided in \cite{AdaptiveNoiseMitigation-2010}.

We set $n_1=20$ and $n_2=10$ as the number of neurons in the first and the second hidden layers, respectively. With three input features and according to Fig.~\ref{fig:DNN}, $\textbf{W}^{(1)}$ is $(20\times3)$ matrix and $\textbf{b}^{(1)}$ is $(20\times1)$ bias vector that connects the input layer to the first hidden layer. After applying the activation function $g^{(1)}$, the matrix $\textbf{W}^{(2)}$ with size $(10\times20)$ and the bias vector $\textbf{b}^{(2)}$ with size $(10\times1)$ will connect the first hidden layer to the second hidden layer. Finally, $\textbf{W}^{(3)}$ is $(1\times10)$ matrix and $\textbf{b}^{(3)}$ is a $(1\times1)$ bias that connects the second hidden layer to the output layer. Since the standard gradient descent from random initialization performs poorly with DNN, the initial values for all parameters is chosen based on Xavier initializer \cite{Xavier_Initializer}. Here, the considered DNN is trained based on the signal model in \eqref{eq:Received_Signal} and noise model in \eqref{eq:Imp_Model}. Specifically, the training set consists of 1000 OFDM symbols with a range of $E_b/N_0$ and SIR that span the operating regions of interested. The samples with different $E_b/N_0$ and SIR values in the training data set is randomly shuffled to remove any trend that may exist.

For a quick reference, the simulation parameters for the considered coded OFDM system in fading channel are listed in Table~\ref{tab:Simulation Parameters}. A total of 1024 subcarriers are used with 672 carrying data, 256 pilot, and 96 null subcarriers. Channel estimation is done based on pilot subcarriers which are equally spaced between 1024 subcarriers. A 10-path fading channel is considered with path arrival times following a Poisson distribution with mean 1 $ms$.  The path amplitudes are Rayleigh distributed with exponentially decreasing average power.
\begin{table}[b]
\begin{center}
\captionsetup{labelfont=sc,labelsep=newline}
\caption{\sc{Simulation Parameters}}
\begin{tabular}{ |l||c| }
\hline
\textbf{Parameters} & \textbf{Values} \\
\hline
\hline
 Bandwidth ($BW$) & 6 kHz \\
 No. of Subcarriers ($N$) & 1024 \\
 Symbol Duration ($T$) & 170.7 ms \\
 Modulation Scheme & QPSK\\
 Channel Length ($L$) & 10\\
 Convolution Code Rate ($CR$) & 1/2 \\
 Code Constraint Length & 7 \\
 Generator Polynomial & [171,133] \\
 Learning Rate ($\eta$)& 0.01\\
 Regularization Hyper Parameter ($\lambda$) & 0.1 \\
 No. of Samples ($n$) & 5 \\

 \hline
\end{tabular}
\label{tab:Simulation Parameters}
\end{center}
\end{table}

\begin{figure}[t]
    \centering
    \begin{subfigure}[b]{0.5\textwidth}
        \includegraphics[width=\textwidth,height=60mm]{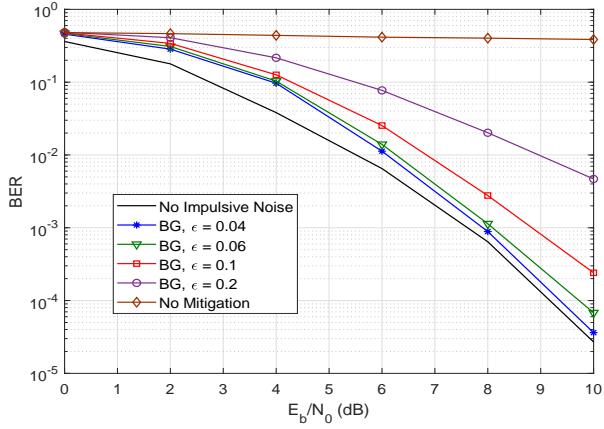}
        \caption{BER in BG noise, SIR = 0 dB.}
        \label{fig:BER_BG}
    \end{subfigure}
    ~ 
    \begin{subfigure}[b]{0.5\textwidth}
        \includegraphics[width=\textwidth,height=60mm]{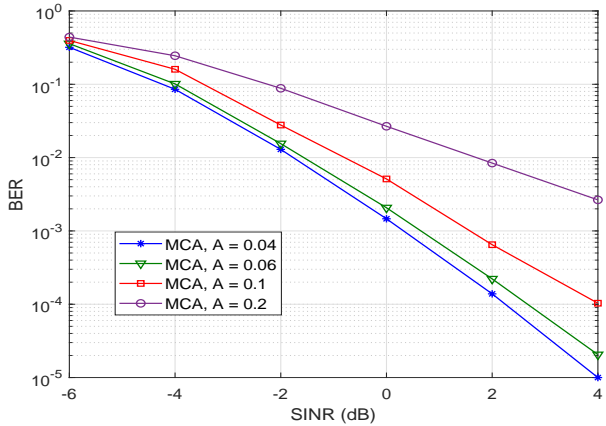}
        \caption{BER in MCA noise, $\Gamma$ = 0.2.}
        \label{fig:BER_MCA}
    \end{subfigure}
    ~ 

    \caption{BER performance of DNN for different model of IN.}
    \label{fig:Importance of DNN}
\end{figure}

\begin{figure}[t]
    \centering
    \begin{subfigure}[b]{0.5\textwidth}
        \includegraphics[width=\textwidth,height=60mm]{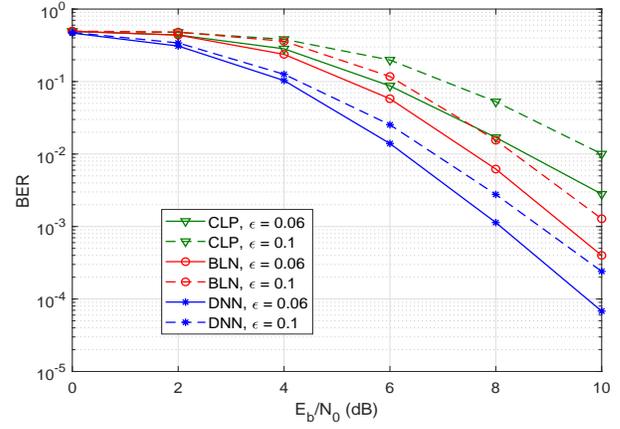}
        \caption{BER comparison in BG noise, SIR = 0 dB.}
        \label{fig:BER_Comp_BG}
    \end{subfigure}
    ~ 
    \begin{subfigure}[b]{0.5\textwidth}
        \includegraphics[width=\textwidth,height=60mm]{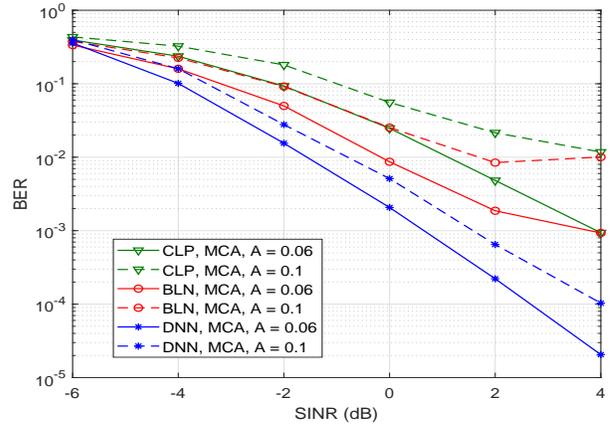}
        \caption{BER comparison in MCA noise, $\Gamma$ = 0.2.}
        \label{fig:BER_Comp_MCA}
    \end{subfigure}
    ~ 

    \caption{BER comparison of DNN, BLN, and CLP for different model of IN.}
    \label{fig:BER Comparison}
\end{figure}

The BER performance of the proposed DNN-based IN mitigation approach under two different test settings (i) BG noise with SIR = 0 dB, and (ii) MCA with $\Gamma$ = 0.2 and $J${=}10 are shown in Fig.~\ref{fig:BER_BG} and Fig.~\ref{fig:BER_MCA}, respectively. As expected the BER performance will degrade with increase in the frequency of IN occurrence. Fig.~\ref{fig:BER Comparison} compares the BER performance of the DNN with blanking (BLN) and clipping (CLP) for different IN models in various levels of impulsivity. From Fig.~\ref{fig:BER Comparison}, it is evident that DNN outperforms both blanking and clipping in all scenarios of both BG and MCA noise models with gains close to 2 dB at BER of $10^{-3}$. Fig.~\ref{fig:BER_Comp_MCA} shows that at high SINR (signal to impulsive plus thermal noise ratio), blanking and clipping are very vulnerable as the level of peakedness decreases and it is difficult to find a proper threshold to distinguish between desired and contaminated signals. On the other hand, a well trained DNN can handle the IN detection process even when the signal and IN peakedness is low. Although, the performance loss of DNN with increase in the frequency of IN occurrence is noticeable, it still outperforms other approaches in all scenarios.

\begin{figure}
\centering
\includegraphics[width=.5\textwidth,height=60mm]{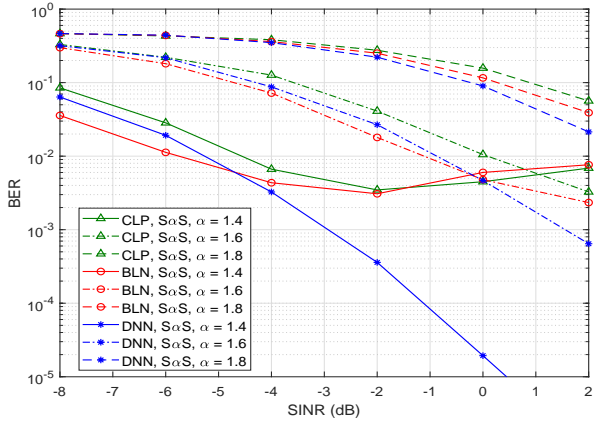}
\caption{BER comparison of DNN, BLN, and CLP in S$\alpha$S noise. $\beta$ = 0, $\gamma {=} 1$, $\mu$ {=} 0.}
\label{fig:BER_Comp_SaS}
\end{figure}

\begin{figure}
\centering
\includegraphics[width=.5\textwidth,height=60mm]{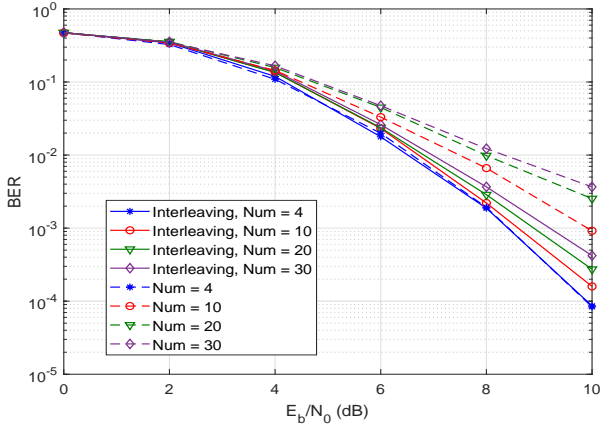}
\caption{BER performance of DNN in bursty IN, SIR = 0 dB, $\epsilon$ = 0.06.}
\label{fig:BER_Comp_Burst}
\end{figure}

Fig.~\ref{fig:BER_Comp_SaS} illustrates the robustness of the proposed DNN approach under IN model mismatch. Although the proposed DNN is trained based on the noise model in \eqref{eq:Imp_Model}, the DNN-based method is the most robust technique relative to blanking and clipping in S$\alpha$S noise model. The performance degradation in blanking and clipping comes from the fact that the threshold calculation is performed based on Gaussian mixture assumption for the received signal which does not hold in this scenario. Fig.~\ref{fig:BER_Comp_Burst} also investigates the BER performance of the considered DNN-based method in bursty IN environment when a time domain interleaver is included in the receiver. In Fig.~\ref{fig:BER_Comp_Burst}, the parameter Num denotes the number of consecutive contaminated samples by IN. As shown in Fig.~\ref{fig:BER_Comp_Burst}, the DNN is able to find the IN instances while the level of burstiness can be alleviated by time domain interleaving. From Fig.~\ref{fig:BER_Comp_Burst} it is obvious that the best performance is achieved when the duration of IN is short.

\section{Conclusions}\label{sec:Conclusion}

In this work, a deep neural network (DNN) is proposed to determine if a received sample is contaminated with impulsive noise (IN) or not in an OFDM-based communication system. The Rank-Ordered Absolute Differences (ROAD) along with median deviations filter is used as input features for the DNN. In the next stage, a nonlinear preprocessor such as blanking is used to suppress the effect of IN in corrupted samples. Simulation results show that the DNN-based approach offers significant improvement in the BER performance in the presence of strong impulsive component. Moreover, the DNN-based IN mitigation outperforms other conventional threshold-based outlier mitigation methods such as blanking and clipping with providing lower BER in IN environments. We also show that DNN-based approach is robust to IN model mismatches and can effectively deal with bursty IN when the receiver includes time domain interleaving. To extend this work one can exploit reinforcement learning to accomplish the impulsive noise mitigation.

\bibliographystyle{IEEEtran}

\bibliography{IEEEabrv,Reference}

\end{document}